\documentclass{article}

\usepackage{arxiv}

\usepackage[utf8]{inputenc} 
\usepackage[T1]{fontenc}    
\usepackage{hyperref}       
\usepackage{url}            
\usepackage{booktabs}       
\usepackage{amsfonts}       
\usepackage{nicefrac}       
\usepackage{microtype}      
\usepackage{lipsum}		
\usepackage{graphicx}
\usepackage{natbib}
\usepackage{doi}
\usepackage{amsmath}
\usepackage{amssymb}
\usepackage{xcolor}

\title{Combining band-frequency separation and deep neural networks for optoacoustic imaging}

\author{ \href{https://orcid.org/0000-0003-0890-0516}{\includegraphics[scale=0.06]{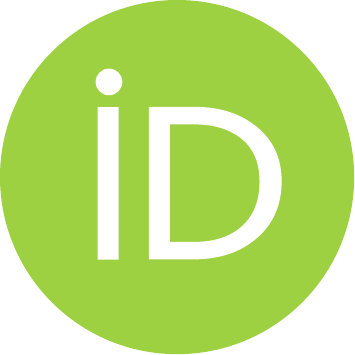}\hspace{1mm}Mart\'in G.~Gonz\'alez}\thanks{Corresponding author. Paper submitted to Optics and Lasers in Engineering.} \\
	Universidad de Buenos Aires and CONICET\\
	Facultad de Ingenier\'ia\\
	Buenos Aires, Argentina \\
	\texttt{mggonza@fi.uba.ar} \\
	\And
	\href{https://orcid.org/0000-0001-9180-7595}{\includegraphics[scale=0.06]{orcid.pdf}\hspace{1mm}Matias Vera} \\
	Universidad de Buenos Aires and CONICET\\
	Facultad de Ingenier\'ia\\
	Buenos Aires, Argentina \\
	\texttt{mvera@fi.uba.ar} \\
    \And
	\href{https://orcid.org/0000-0002-5578-0521}{\includegraphics[scale=0.06]{orcid.pdf}\hspace{1mm}Leonardo J. ~Rey Vega} \\
	Universidad de Buenos Aires and CONICET\\
	Facultad de Ingenier\'ia\\
	Buenos Aires, Argentina \\
	\texttt{lrey@fi.uba.ar} \\
}

\date{}

\renewcommand{\headeright}{}
\renewcommand{\undertitle}{}
\renewcommand{\shorttitle}{}

\hypersetup{
pdftitle={Combining band-frequency separation and deep neural networks for optoacoustic imaging},
pdfsubject={eess.IV, cs.AI},
pdfauthor={Martin G.~Gonzalez, Matias A.~Vera, Leonardo J. ~Rey Vega},
pdfkeywords={Tomography, Optoacoustic, Deep Learning, UNet},
}

\begin{document}
\maketitle

\begin{abstract}
	In this paper we consider the problem of image reconstruction in optoacoustic tomography. In particular, we devise a deep neural architecture that can explicitly take into account the band-frequency information contained in the sinogram. This is  accomplished by two means. First,  we jointly use a linear filtered back-projection method and a fully dense UNet for the generation of the images corresponding to each one of the frequency bands considered in the separation. Secondly, in order to train the model, we introduce a special loss function consisting of three terms: (i) a separating frequency bands term; (ii) a sinogram-based consistency term and (iii) a term that directly measures the quality of image reconstruction and which takes advantage of the presence of ground-truth images present in training dataset. Numerical experiments show that the proposed model, which can be easily trainable by standard optimization methods, presents an excellent generalization performance quantified by a number of metrics commonly used in practice. Also, in the testing phase, our solution has a comparable (in some cases lower) computational complexity, which is a desirable feature for real-time implementation of optoacoustic imaging. 
\end{abstract}

%
%
%
%


\section{Introduction}
\label{sec:intro}

Optoacoustic tomography (OAT) is an imaging technique based on the optoacoustic (OA) effect. By using laser excitation and ultrasonic detectors, OAT takes advantage of the high contrast imaging present in purely optical techniques while maintaining the great resolution given by ultrasonic detection \cite{xu2006}. 
The illumination of biological tissue with non-ionizing short laser pulses leads to a rapid increase in temperature and to the formation of pressure waves due to thermoelastic expansion of the sample under study. Acoustic waves propagates through the sample and are sensed by wideband ultrasonic transducers, typically placed around the sample \cite{paltauf2017}, \cite{Awasthi_Jain_Kalva_Pramanik_Yalavarthy_2020}. The detected signals (referred to as the sinogram), which contains valuable information about the sample, are then fed to specialized numerical algorithms in order to recover the initial pressure induced by laser light absorption. As optical absorption is linked with several physiological properties, among others oxygen saturation and hemoglobin concentration, several diagnostic applications are well-suited for this technique \cite{Hauptmann_Cox_2020}, \cite{tian2020}.

Besides the problem of implementing a proper OA system capable of generating the exciting optic signal and detecting the acoustic signals generated by the sample under study, one major challenge is the design of the algorithms that are responsible for processing the detected signals and deliver the initial pressure profile generated in the sample after laser illumination \cite{lutzwieler2013}. There exists several approaches for the reconstruction of the pressure distribution induced by laser illumination, that can be classified as analytical or algebraic \cite{rosenthal2013}. Analytic reconstruction techniques, such as the Back-Projection (BP) \cite{xu2005} algorithms, are characterized by inverting the exact forward acoustic operator by means of the analytical inversion of the mathematical equations. For example, there exist well-known analytic results for usual geometries \cite{Minghua_Xu_Wang_2002}, \cite{Minghua_2003} and that includes other effects as form factors of the acoustic sensors employed \cite{Burgholzer_Bauer-Marschallinger_Grün_Haltmeier_Paltauf_2007}. However, analytic techniques fall short in considering important effects such as the unavoidable presence of noise and modelling mismatches and typically require a large amount of data to provide an accurate image reconstruction. Algebraic reconstruction techniques, on the other hand, consider the discretization or approximation of the underlying physical model (direct or inverse). Algebraic approaches, such as the model-based-matrix (MBM) algorithm \cite{Rosenthal_Razansky_Ntziachristos_2010}, have sound theoretical foundations, are well-studied and are widely recognized as baseline benchmarks against which new reconstruction methods techniques are compared. Besides some standard statistical assumptions with respect to the measurement noise that affect the acquired signals, these techniques also makes a full use of the underlying physical principles of the application. 
They also are highly versatile as they easily allow the inclusion of a different number of constraints for the reconstruction problem. For example, it is common to include Tikhonov regularization and positivity constraints \cite{Ding_2015}, total variation constraints \cite{Huang_Wang_Nie_Wang_Anastasio_2013} or $L_1$ regularization terms that promote sparsity features in the reconstructed images or in the sinogram  \cite{Provost_Lesage_2009},\cite{Haltmeier_Sandbichler_Berer_Bauer-Marschallinger_Burgholzer_Nguyen_2018}, \cite{Betcke_Cox_Huynh_Zhang_Beard_Arridge_2017}. As another application of this versatility, in \cite{Longo_Justel_Ntziachristos_2022}, an explicit frequency disentanglement of the broadband measured acoustic signals is considered, which introduces interesting quality improvements in the image reconstruction. As an important downside, algebraic methods are typically iterative in nature, requiring important memory requirements and computational load, which translate in large processing times. 

We can say that algebraic and analytic techniques are \emph{model-guided} approaches for the reconstruction problem. However, their formulation is mainly based on somewhat ideal measurement settings, which are not fully representative of a real situation. These mismatch modelling issues typically introduce artifacts in the reconstructed images. Issues as the sensors shape, filtering effects in the acquisition chain, uncertainty in the speed of sound of the sample and/or in the sensor positions are among the most common mismatches \cite{Sahlstrom2020}. One possible approach to this issue is the inclusion of specially designed crafted matrix operators or the application of Bayesian techniques to make the algebraic reconstruction method aware of those effects. Another way is to combine the \emph{model-guided} approach described above, which capture the main characteristic of the underlying physical problem, with a powerful \emph{data-driven} approach. Data-driven will make use of the information contained in a dataset of numerically simulated measurements and/or true experimental data in order to \emph{learn and correct} those aspects no contained in the ideal physical model. Deep learning architectures are specially well-suited for this \cite{LeCun_Bengio_Hinton_2015}. Specifically, there is growing interest in including \emph{model-guided} information along deep neural nets to better exploit the expressive capacities of these structures with the goal of improving performance in the task at hand, and/or decreasing the need of large amounts of training data\cite{Shlezinger_Whang_Eldar_Dimakis_2020}. Some applications and insights of this approach to the problem of image reconstruction in OAT can be consulted in \cite{Hauptmann_Cox_2020}. Given the expressive capacity of deep neural nets, if the selected deep structure is finely tuned, reconstruction artifacts induced by modelling mismatches can be significantly reduced. Moreover, once the deep net is trained, the computational load required for processing a given sinogram to obtain the original image is considerable lower than the one corresponding to a typical algebraic approach, which has an enormous impact in the reconstruction times and computational resources needed for real-time OAT reconstruction. Of course, this comes at the expense of an initial numerically intensive training procedure with a sufficiently large database.

In this paper, we study a reconstruction problem that take into account a frequency disentanglement of the measured sinogram, similar to one proposed in \cite{Longo_Justel_Ntziachristos_2022}, in order to better exploit the wideband nature of the OA signals. However, instead of relying in a fully algebraic model-guided approach for the image reconstruction, we consider an hybrid \emph{model-guided-data-driven} method for the proper training of a deep neural network. To achieve this, we employ a convolutional neural network with an appropriate cost function, that allows to obtain excellent performance (measured with respect to several qualitative and quantitative metrics) in the image reconstruction task. 

The paper is organized as follows. In Section \ref{sec:reconstruction} we summarize the major mathematical details of the acoustic reconstruction problem in OAT. In Section \ref{sec:method} we detail our proposal. In Section \ref{sec:results} the merits of our proposal are evaluated numerically and experimentally. Finally, in Section \ref{sec:conclu}, some concluding remarks are discussed.   

\section{The reconstruction problem}
\label{sec:reconstruction}

\subsection{The forward problem}
\label{subsec:met_forward}

It is well-known that  after the excitation of a biological sample by an electromagnetic pulse $\delta(t)$, the acoustic pressure $p(\mathbf{r},t)$ at position $\mathbf{r} \in\mathbb{R}^3$ and time $t$, satisfies \cite{Wang_Wu_2007}:

\begin{equation}
\left(\frac{\partial^2}{\partial t^2} - v_s^2 \, \nabla^2 \right) p(\mathbf{r},t) = 0
\label{eq:waveeq}
\end{equation}

\noindent with the initial conditions,

\begin{equation}
p(\mathbf{r},0) = p_0(\mathbf{r})\, \text{,} \quad \left(\partial p /\partial t\right)(\mathbf{r},0)= 0 
\label{eq:waveeq_cond_ini}
\end{equation}

\noindent where $p_0(\mathbf{r})$ is the initial OA pressure and $v_s$ represents the speed of sound in the medium, which is assumed acoustically non-absorbing and homogeneous. Under the usual hypothesis of thermal and acoustic confinement \cite{Kruger_Liu_Fang_Appledorn_1995}, that is, when the laser pulse duration is short enough such that the heat conduction and acoustic propagation into neighboring regions of the illuminated region can be neglected, the initially induced pressure $p_0(\mathbf{r})$ is proportional to the total absorbed optical energy density. Using Green's function formalism, the pressure received by an ideal point-detector at position $\mathbf{r_d}$ can be written as:

\begin{equation}
p_d(\mathbf{r_d},t)=\frac{1}{4\pi\,v_s^2} \frac{\partial}{\partial t}\iiint_{V} \, p_0(\mathbf{r}) \frac{\delta\left(t-|\mathbf{r_d}-\mathbf{r}|/v_s\right)}{|\mathbf{r_d}-\mathbf{r}|} d^3\mathbf{r}
\label{eq:fo_time}
\end{equation}

The goal of the OAT inverse problem is to reconstruct $p_0(\mathbf{r})$ from the signals $p_d(\mathbf{r_d},t)$ measured at various positions $\mathbf{r_d}$, which are typically in a surface $S$ that contains the volume of interest \cite{lutzwieler2013}.

\subsection{The inverse problem}
\label{subsec:met_inverse}

Possibly, the most popular reconstruction approach in OAT are BP type algorithms, due to their simple implementation and applicability to a lot of practical imaging scenarios \cite{rosenthal2013}. One of the most important formulations of the BP approach is the universal back-projection algorithm \cite{xu2005}. In a homogeneous medium with a constant $v_s$, the universal BP formula directly links $p_0(\mathbf{r})$ to $p_d(\mathbf{r_d},t)$ on the detection surface $S$ that encloses the OA source \cite{xu2005}:

\begin{equation}
p_0(\mathbf{r})=\int_{\Omega_s} b\left(\mathbf{r_d},t=|\mathbf{r_d}-\mathbf{r}|/v_s\right) \, \frac{d\Omega_s}{\Omega_s}
\label{eq:ubp}
\end{equation}

\noindent where $b(\mathbf{r_d},t)=2\,p_d(\mathbf{r_d},t)-2\,t\,\partial p_d(\mathbf{r_d},t)/\partial t$ is the BP term related to the measurement at position $\mathbf{r_d}$, $\Omega_s$ is the solid angle of the whole surface $S$ with respect to the reconstruction point inside $S$, $d\Omega_s = dS \,\cos \theta_s/|\mathbf{r_d}-\mathbf{r}|$ and $\theta_s$ denotes the angle between the outwards pointing unit normal of $S$ and $(\mathbf{r_d}-\mathbf{r})$. The above formula provides exact inversion for several interesting geometries (e.g. cylindrical, planar and spherical). However, it assumes that the detectors are point one with no bandwidth limitations and isotropic angular response \cite{rosenthal2013}. However, in practice, the transducers are extended, have a limited bandwidth and their spatial response is not constant. In these non-ideal imaging scenarios, (\ref{eq:ubp}) significantly deviate from reality, generating imaging artifacts and distorted images. Moreover, (\ref{eq:ubp}) assumes that the detected signals are not noisy, which is not the typical case in practice.

\begin{figure}
  \centering
   \hspace*{2.05cm}
  \includegraphics[width=9cm]{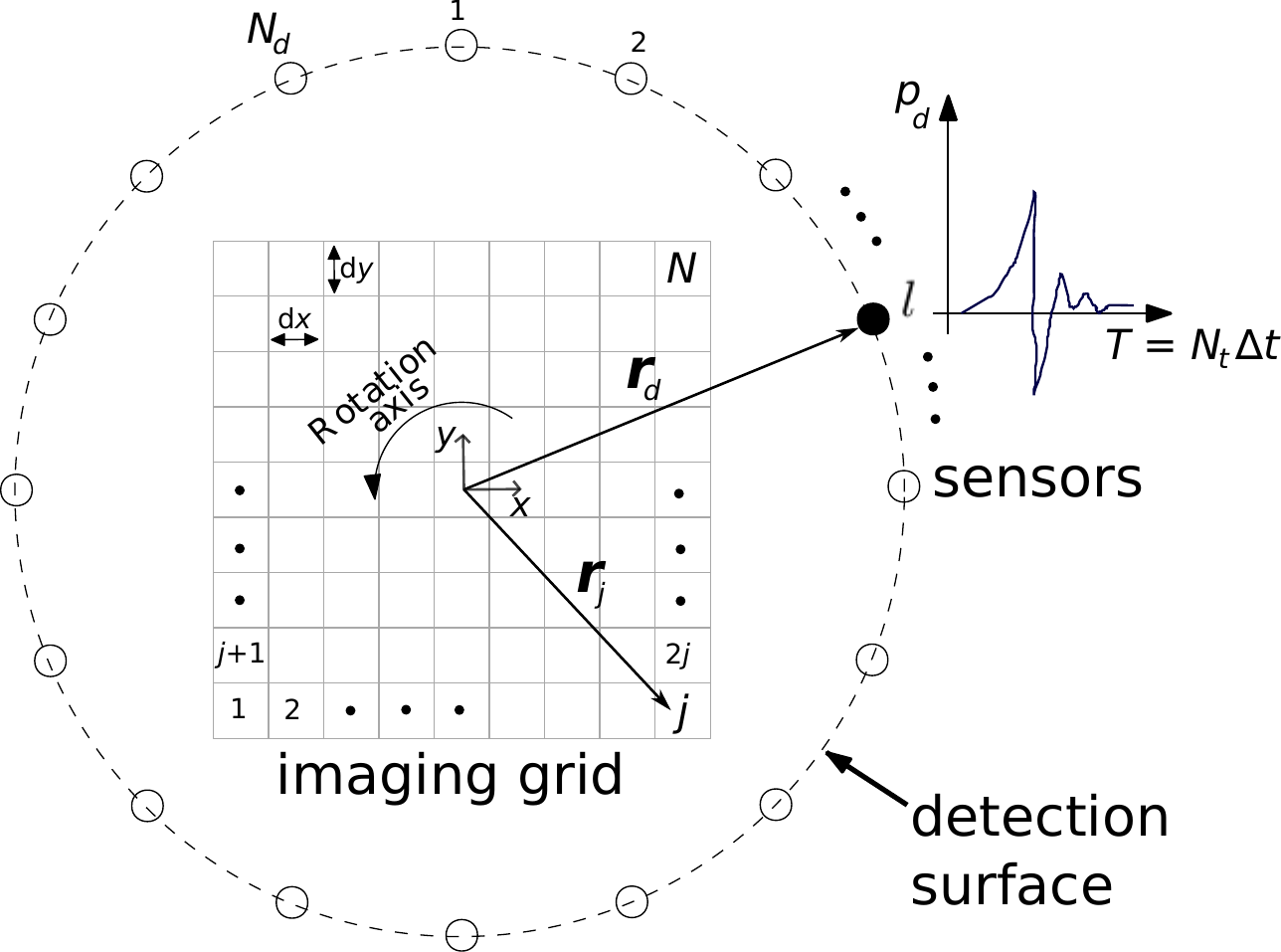}
    \caption{Schematic of the OAT imaging setup studied in this work. A number of $N_d$ ultrasonic sensors are uniformly distributed around the sample which is divided in an imaging grid of $N$ pixels. The signals at each sensor are observed during a time window of length $T$ and sampled with sampling period $\Delta t$ to obtain $N_t$ samples for each sensor.}
\label{fig:setup}
\end{figure}

A different approach to the reconstruction problem is given by a MBM algorithm \cite{Rosenthal_Razansky_Ntziachristos_2010}. In this technique, the forward solution in (\ref{eq:fo_time}) is discretized. As a result a matrix equation is obtained, which is used for solving the inverse problem. One of the advantages of this approach is that any linear effect in the system may be easily considered (e.g. sensor form factors, linear filtering or the spatial response of the sensors): 

\begin{equation}
\mathbf{p_d}=\mathbf{A} \, \mathbf{p_0}
\label{eq:mbt}
\end{equation}

\noindent where $\mathbf{p_d} \in\mathbb{R}^{N_d \cdot N_t\times 1}$ is a column vector representing the measured pressures at a set of detector locations $\mathbf{r_d}_l$ ($l=1 \ldots N_d$) and time instants $t_k$ ($k=1 \ldots N_t$); $\mathbf{p_0} \in\mathbb{R}^{N\times 1}$ is a column vector representing the values of the initial acoustic pressure on the imaging region grid; and $\mathbf{A} \in\mathbb{R}^{N_d \cdot N_t\times N}$ is the model matrix. The $j$-th element ($j=1 \ldots N$) in $\mathbf{p_0}$ contains the average value of the initial pressure within a volume element of size $\Delta V$ at position $\mathbf{r}_j$. Once the discrete formulation has been established, the inverse problem is reduced to the algebraic problem of inverting (\ref{eq:mbt}). The matrix $\mathbf{A}$ can be written as the multiplication of two matrices $\mathbf{A^{oa} \, A^s}$ where $\mathbf{A^s}$ represents the response function of the imaging system for an ideal point-like sensor and $\mathbf{A^{oa}}$ is the matrix form of a time derivative operator. The matrix $\mathbf{A^s}$ is defined as\cite{paltauf2018}: 

\begin{equation}
    A^s_{lkj} = \frac{1}{4\pi v_s^2}\frac{\Delta V}{\Delta t^2} \frac{d(t_k,\mathbf{r}_j,\mathbf{r}_{dl})}{|\mathbf{r_d}_l - \mathbf{r}_j|}
\label{eq:Gs1}
\end{equation}

\begin{equation}
    d(t_k,\mathbf{r}_j,\mathbf{r}_{dl}) = \begin{cases}
    1 & \text{si } |t_k - \frac{|\mathbf{r_d}_l - \mathbf{r}_j|}{v_s}| < \Delta t/2  \\
    0 & \text{otherwise} 
  \end{cases}
    \label{eq:Gs2}
\end{equation}

\noindent where $\Delta t$ is the time step at which the signals $p_d(\mathbf{r_d},t)$ are sampled. It is not difficult to see that (\ref{eq:Gs1}) constitutes a discretization of the integrand in (\ref{eq:fo_time}), while (\ref{eq:Gs2}) indicates the time at which the effect of initial pressure at position $\mathbf{r}_j$ is captured by sensor $\mathbf{r_d}_{l}$.
In the case of a ﬁnite-size detector, the spatial impulse response (SIR) of the sensor is taken into account by dividing the area of the sensor into surface elements (treated as point detectors) which are then added up \cite{rosenthal2011, paltauf2018}. A typical OAT imaging setup is shown in Fig. \ref{fig:setup}.

The inversion of \eqref{eq:mbt} is typically done using a quadratic criterion plus a Tikhonov regularization term:
\begin{equation}
    \mathbf{\hat{p}_0} = \min_{\mathbf{p_0}} || \mathbf{A} \, \mathbf{p_0} - \mathbf{p_d} ||^2 + \lambda \, ||\mathbf{p_0}||^2 
    \label{eq:regTik}
\end{equation}

\noindent where $\lambda\geq 0$ is parameter that improves the stability of inverse problem (which is typically ill-conditioned), and it also have beneficial effects when noise is present in the measured signals. Given a fixed $\lambda$, the solution to \ref{eq:regTik} is unique and given by \cite{Provost_Lesage_2009}:

\begin{equation}
\mathbf{\hat{p}_0} = (\mathbf{A}^H\,\mathbf{A} + \lambda \mathbf{I})^{-1} \mathbf{A}^H\,\mathbf{p_d}
\label{eq:solTik}
\end{equation}
\noindent where $\mathbf{I}$ is the identity matrix and $H$ denotes the conjugate transpose operator. The value of the regularization parameter $\lambda$ has a significant effect on the solution and must be carefully chosen. Although (\ref{eq:solTik}) is the exact solution to problem (\ref{eq:regTik}), it is rarely use in practice. It is more efficient, from a computationally point of view, to use a quadratic solver \cite{Paige_Saunders_1982}. Even, in this case, given the large size of the system of (\ref{eq:mbt}), the computational resources needed to reconstruct only one image are important. However, as explained in the introduction, the formalism given by the cost criterion in (\ref{eq:regTik}), allows to easily include several additional and useful constraints in the problem such as positive constraints in the recovered image pixels. For example, in order to explicitly use the usual broadband nature of the sinograms and being able to efficiently extract the full information contained in them, in \cite{Longo_Justel_Ntziachristos_2022} a frequency-band model-based (fbMB) algorithm with explicit soft-priors in the frequency domain is proposed. In more precise terms, the images $x_1,\dots,x_n$ corresponding to the OA signal frequency content in the $1,2,\dots,n^{th}$ bands, in which the full bandwidth of the OA signals is divided, are obtained by solving:
\begin{equation}
    (\mathbf{x}_1^{*},\dots,\mathbf{x}_n^{*})=\arg\min_{(\mathbf{x}_1,\dots,\mathbf{x}_n)\geq 0}\|\mathbf{p_d}-\mathbf{A}(\mathbf{x}_1+\mathbf{x}_2+\dots+\mathbf{x}_n)\|^2+\lambda\|\mathbf{x}_1+\mathbf{x}_2+\dots+\mathbf{x}_n\|^2+\eta\sum_{k=1}^n\mu_k\|\mathbf{F}_k\mathbf{A}\mathbf{x}_k\|^2,
\label{eq:loss_1}
\end{equation}

where $\mathbf{F}_k$ is the matrix form of a band-reject filter for the $k^{th}$ band and $\eta$, $\lambda$ and $\mu_i$ are hyperparameters. 
After recovering  $(\mathbf{x}_1^{*},\dots,\mathbf{x}_n^{*})$, the final reconstruction is computed as $\mathbf{\hat{p}_0}=\sum_{k=1}^n \mathbf{x}_k^{*}$.
It is shown that fbMB provides high contrast and accurate reconstructed images. However, the computational load for reconstructing a given image is even higher than (typically a $n^{th}$ fold increase in computational load) the problem in (\ref{eq:regTik}).

\section{Proposed method}
\label{sec:method}
\subsection{Network architecture and loss function}

In this section, and recognizing the value in the fbMB solution of (\ref{eq:loss_1}), we consider using the properties and benefits of frequency disentanglement of the measured sinogram along with a specially suited deep neural architecture. Deep neural networks has been extensively used in different tasks in biomedical imaging \cite{Kim_Yun_Cho_Shin_Jang_Bae_Kim_2019}. Specific implementations for image reconstruction in OAT has received considerable attention in the recent years (see \cite{Hauptmann_Cox_2020},\cite{Awasthi_Jain_Kalva_Pramanik_Yalavarthy_2020} and the references therein). Assuming that a dataset of numerically simulated measurements and/or true experimental data is available, a \emph{data-driven} approach that will make use of the available data to \emph{learn} (and possibly \emph{correct} if some mismatch is present in the used model) the physics mechanics of the problem at hand, can be employed. However, the expressive power to the deep architecture can be biased at training time using information of the OAT physics dynamics. The more direct piece of information that can be easily used is the discretized forward operator in (\ref{eq:mbt}), which efficiently captures the essentials of the problem at hand. 

We will consider a neural network $G(\mathbf{x};\theta):\mathbb{R}^N\rightarrow\mathbb{R}^{N\times n}$, that take as inputs images (or initial pressure profiles) and deliver $n$ images of the same dimension (one for each of the $n$ disentangled frequency bands as in (\ref{eq:loss_1})). $\theta$ are the parameters of the architecture that should be learned during training using an appropriate database. The image $\mathbf{x}$ that the network $G$ will take as input is an initial image reconstruction delivered by an easy to implement method. For example, a simple and efficient choice is to consider the image delivered by:
\begin{equation}
\mathbf{x}=\mathbf{A}^{T}\mathbf{p_d},
\label{eq:adjoint}    
\end{equation}
where $\mathbf{A}^T$ is the adjoint of $\mathbf{A}$. This initial image reconstruction is basically a linear filtered BP (LBP) method \cite{Hauptmann_Cox_2020}, that in general do not give the best results in image reconstruction but is efficient and numerically robust \cite{Hoelen_Mul_2000}. The main idea is that the neural net $G$ takes as input $\mathbf{A}^{T}\mathbf{p_d}$, and using its expressive capacity leveraged by a training database, could deliver a better image quality, learning to correct artifacts, modelling mismatchs and other impairments. 

In order to properly find the parameters $\theta$ of the neural net $G$, we consider the following loss function:
\begin{equation}
    l(\mathbf{p_d},\mathbf{p_0};\theta)=\Big\|\mathbf{p_d}-\mathbf{A}\sum_{k=1}^n\left[G(\mathbf{A}^{T}\mathbf{p_d};\theta)\right]_k\Big\|^2+\eta\sum_{k=1}^n\mu_k \Big\|\mathbf{F}_k\mathbf{A}\left[G(\mathbf{A}^{T}\mathbf{p_d};\theta)\right]_k\Big\|^2+\eta_I\Big\|\mathbf{p_0}-\sum_{k=1}^n \left[G(\mathbf{A}^{T}\mathbf{p_d};\theta)\right]_k\Big\|^2,
\label{eq:loss_2}
\end{equation}

\noindent where $\mathbf{x}_k\equiv\left[G(\mathbf{A}^{T}\mathbf{p_d};\theta)\right]_k$ denotes the image reconstructed by $G$ and corresponding to the $k^{th}$ frequency band of the sinogram $\mathbf{p_d}$. Clearly, the total image reconstructed is given by $\sum_{k=1}^n\left[G(\mathbf{A}^{T}\mathbf{p_d};\theta)\right]_k$. Some comments regarding the loss function follows:
\begin{enumerate}
    \item The first term in (\ref{eq:loss_2}) is a data-consistency term. Considering the image recovered by the network $G$, this image is processed by the forward operator $\mathbf{A}$ in order to check that the reconstructed image is consistent with the full measured sinogram information given by $\mathbf{p_d}$.
    \item The second term is responsible of the frequency disentaglement of the reconstructed $k^{th}$ frequency band sinogram $\mathbf{A}\left[G(\mathbf{A}^{T}\mathbf{p_d};\theta)\right]_k$ (whose consistency is check by the first term). This term (and the first one) are conceptually similar to the first and third term in (\ref{eq:loss_1}).
    \item The third term consider the quality of the reconstructed image with respect to the ground-truth $\mathbf{p_0}$. Although several possible metrics could be considered, we choose the simple mean-square error to measure the quality of reconstruction. Clearly, this term has not equivalent in the original formulation in (\ref{eq:loss_1}).
    \item The positive hyperparameters $\eta$, $\eta_I$ and $\left\{\mu_k\right\}_{k=1}^n$ weight the different terms in the loss function and are chosen using a validation test at training time.
\end{enumerate}
Assuming a $M$-length training database $\left\{\mathbf{p_d}^i,\mathbf{p_0}^i\right\}_{i=1}^M$,  is available, where $\mathbf{p_d}^i$ are the measured sinograms and $\mathbf{p_0}^i$ are the ground-truth images (or initial pressure distributions) the parameters $\theta$ are obtained minimizing the functional $\frac{1}{N}\sum_{i=1}^M \ell(\mathbf{p_d}^i,\mathbf{p_0}^i;\theta)$ using a backpropagation procedure.

\begin{figure}
  \centering
  \includegraphics[width=16cm]{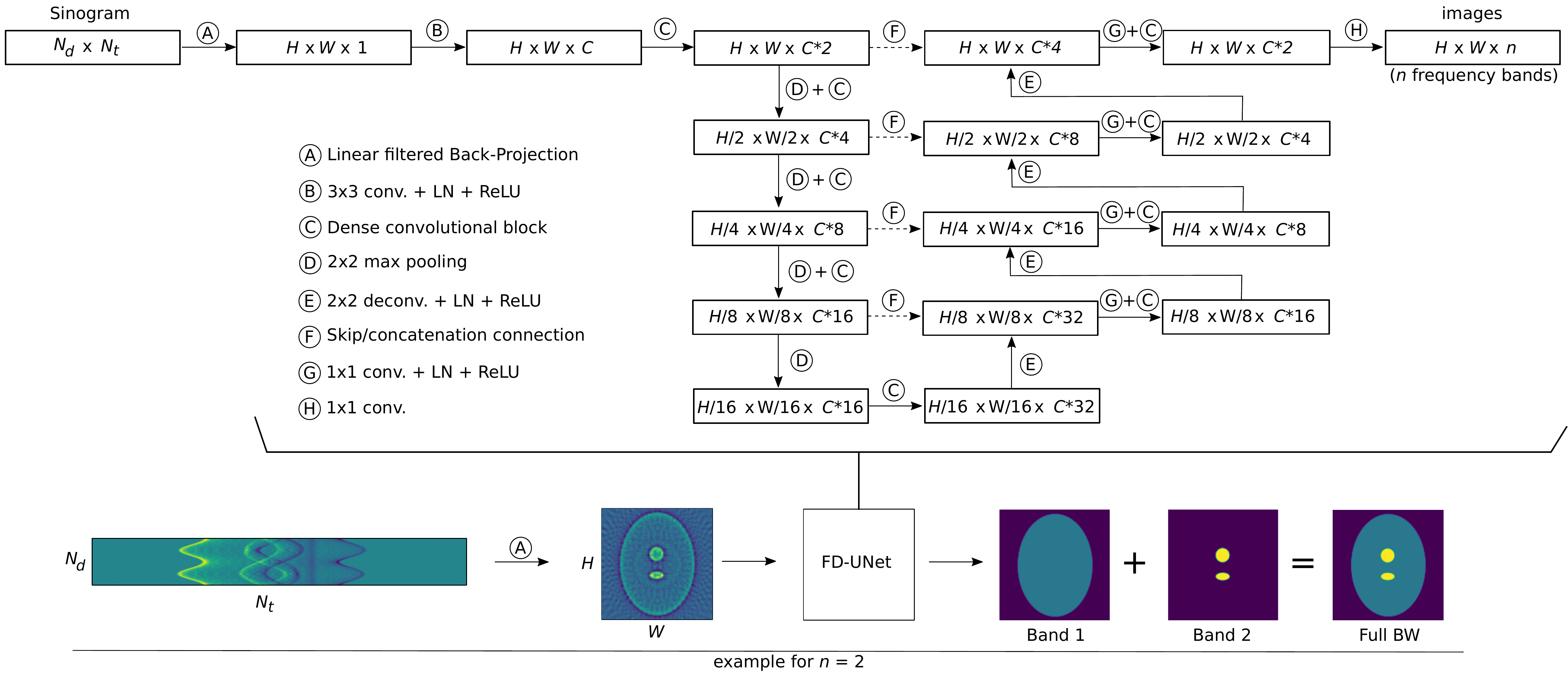}
    \caption{FD-UNet architecture. Common parameters used in our experiments are: $N_d=32, N_t=1024, H=W=128, C=32$ and $n=2$. Schematically the sinogram input, the initial image fed to the network (LBP reconstruction) and the images corresponding to each frequency band and its aggregation which should be the full reconstructed image are also shown.}
\label{fig:fdunet}
\end{figure}

\subsection{Implementation details}
\label{subsec:details}
As architecture for the neural net $G$ we consider a UNet architecture \cite{Ronneberger_Fischer_Brox_2015}, which is basically a multi-scale convolutional autoencoder using a residual connection between input and output and skip connections that connect encoder and decoder at each scale, providing among other things numerical stability during the backpropagation training. Moreover, the different scales at the encoder and decoder include dense connectivity, allowing to a better information flow through the network and robustness against learning redundant features. The use of Fully-Dense UNets (FD-UNet) in OAT is promising, delivering excellent restoration results \cite{guan2020} and artifacts suppression. In Fig. \ref{fig:fdunet} there is a schematic representation of the used architecture, with the common parameters used in our experiments. In our implementation, layer normalization (LN) is used to speed up the training procedure. The implementation code can be found in \url{https://github.com/mggonza/BFSNNOAI}.

Related to the second term of the loss function in (\ref{eq:loss_2}), the choice of band-reject filter matrices $\mathbf{F}_k$, with $k=1,\dots,n$ is done following the recommendations in \cite{Longo_Justel_Ntziachristos_2022} (using fourth-order Butterworth band-pass filters with $BW_{\%}=1.6$). The only difference is that each $\mathbf{F}_k$ is constructed in such a way that the filtering is done in a zero-phase fashion, filtering the $k^{th}$ band sinogram $\mathbf{A}\left[G(\mathbf{A}^{T}\mathbf{p_d};\theta)\right]_k$ in the forward and reverse directions. This improves the frequency separation and avoids the introduction of any phase distortion by the filtering process. 

 In order to train our model we considered a dataset containing synthetic (numbers and letters) and experimental retinal vasculature phantoms from a public database \cite{drive2020}. For generating the corresponding sinograms that are used as input in our method (and the other tested for comparison in Section \ref{sec:results}), we used (\ref{eq:Gs1}) considering the setup of Fig. \ref{fig:1}, which according to our extensive numerical simulations gives very similar results to using the popular k-Wave toolbox \cite{kwave}, but requires significantly less computation time. In order to allow for typical effects that can be usually found in real-world settings, we added random perturbations around the nominal discretized forward operator for the generation of the sinograms for each image in the database. The perturbed quantities were the speed of sound ($1485 \pm 10)$ m/s, the sensor positions (0.1\% perturbation around nominal positions) and the aggregation of Gaussian measurement noise with different SNRs (20-80 dB). As the transpose of the discretized forward operator $\mathbf{A}$ at the deep network input is always the same (the nominal and unperturbed one), the deep network should also learn to be robust to these small variations in the forward operator.

\section{Results}
\label{sec:results}


To test the method we used a setting similar to the one in Fig. \ref{fig:setup} with images of $128\times 128$ pixels (pixel size of 50 $\mu$m) and 32 sensors uniformly distributed around the sample. The sampling frecuency was set to 78.8 MHz. The sensors had a random uncertainty of position of 0.1\%.  Each sensor signal had a duration of 13 $\mu$s (1024 samples) and white Gaussian noise was added with variable variance levels leading to sinogram measurement SNR values between 20 dB and 80 dB. First, we considered $n=2$ frequency bands, where the band-pass Butterworth filters used to implement the band-reject filters were designed with bands $[0.18 \mbox{ MHz}-1.65 \mbox{ MHz}]$  and  $[1.65 \mbox{ MHz}-15 \mbox{  MHz}]$, respectively. The first band is related to the resolution of details in the range of 0.9 - 8 mm. This  range is related with the full size of the images $(6.4 \times 6.4) \text{ mm}^2$ and some large structures easily spotted in the images (large vessels). The second band allows resolving details in the range of 0.1-0.9 mm. Notice that the lower limit of 0.1 mm is in line with pixel size of 50 $\mu$m. The FD-UNet has the architecture shown in Fig. \ref{fig:fdunet} with $n=2$, presenting at its output low and high frequency images $x_1$ and $x_2$.  The number of phantoms\footnote{In our tests, we have considered the use of even more images for training (i.e. up to 250000). However, using a number of training examples above to 10000 did not show any significant improvements with respect to ones reported here.} used for training were 10000 and 2000 were used for validation and selection of hyperparameters, e.g. $\eta$ and $\eta_I$. In order to test the optimized model, we considered 600 images not contained in the training database. The hyperparameters in the cost function (\ref{eq:loss_2}) were $\eta=0.01$, $\eta_I=1$ and $\mu_1=\mu_2=0.5$. We used ADAM optimization \cite{Kingma2015AdamAM} with parameters $\beta_1=0.9$ and $\beta_2=0.999$. The initial learning rate, number of epochs and the batch size were set to $10^{-4}$, 100 and 2, respectively. The total training time was of approximately 14 hours in a computer with CPU Intel i7-9700F, 64 GB of RAM and a GPU RTX 2080 with 8 GB of memory.

In Table \ref{table:1} we show the average generalization performance of our approach for different popular metrics: Structural Similarity Index (SSIM), Pearson Correlation (PC), Root Mean Square Error (RMSE) and Peak Signal to Noise Ratio (PSNR). In the same table, it is also presented the performance over the testing set of other popular methods in the literature (Delay and Sum (DAS) and LBP), including fbMB from  \cite{Longo_Justel_Ntziachristos_2022}. Moreover, we considered an optimized FD-UNet architecture similar to the one in \cite{guan2020} and trained with a loss function as in (\ref{eq:loss_2}) but without the first and second terms ($\eta=0$) which are responsible of emphasizing the frequency disentanglement and sinogram-based consistency. We see that the proposed architecture has the best performance for all quantitative metrics considered. In particular, for the usual RMSE metric the average performance gain with respect to the second best (the simple FD-UNet architecture) is 75 \% approximately. With respect to the third best method, which is fbMB, the performance gain is almost 145 \%. Also, the standard deviation of the performance over the testing set for the proposed method is the best among all quantitative measures and the three best performing methods, which gives some idea of the methods precision over the testing set. The results show, uniformly, over the four quantitative metrics employed, that the use of the frequency disentanglement proposed in \cite{Longo_Justel_Ntziachristos_2022} and the expressive power of an FD-UNet, are both important to achieve the best performance. Finally, we should point out that the popular DAS and LBP methods present the worst performance indices. 

\begin{table}
    \centering \caption{Performance (mean value and standard deviation) over the testing set}
    \begin{tabular}{ccccc}
        \hline
        Method & SSIM & PC & RMSE & PSNR \\
        \hline
        Proposed & \textbf{0.879} $\pm$ 0.103 & \textbf{0.965} $\pm$ 0.043 & \textbf{0.047} $\pm$ 0.022 & \textbf{27.528} $\pm$ 4.228 \\
        FD-UNet  & 0.783 $\pm$ 0.154 & 0.941 $\pm$ 0.069 & 0.083 $\pm$ 0.062 & 23.670 $\pm$ 5.966 \\
        fb-MB  & 0.606 $\pm$ 0.181 & 0.909 $\pm$ 0.100 & 0.115 $\pm$ 0.061 & 20.007 $\pm$ 4.597 \\
        LBP  & 0.080 $\pm$ 0.039 & 0.537 $\pm$ 0.137 & 0.349 $\pm$ 0.121 & 9.695 $\pm$ 3.162 \\
        DAS  & 0.023 $\pm$ 0.020 & 0.283 $\pm$ 0.163 & 0.522 $\pm$ 0.193 & 6.480 $\pm$ 3.513 \\
        \hline
   \end{tabular}
\label{table:1}
\end{table}

\begin{figure}[t]
  \centering
  \includegraphics[width=12.5cm]{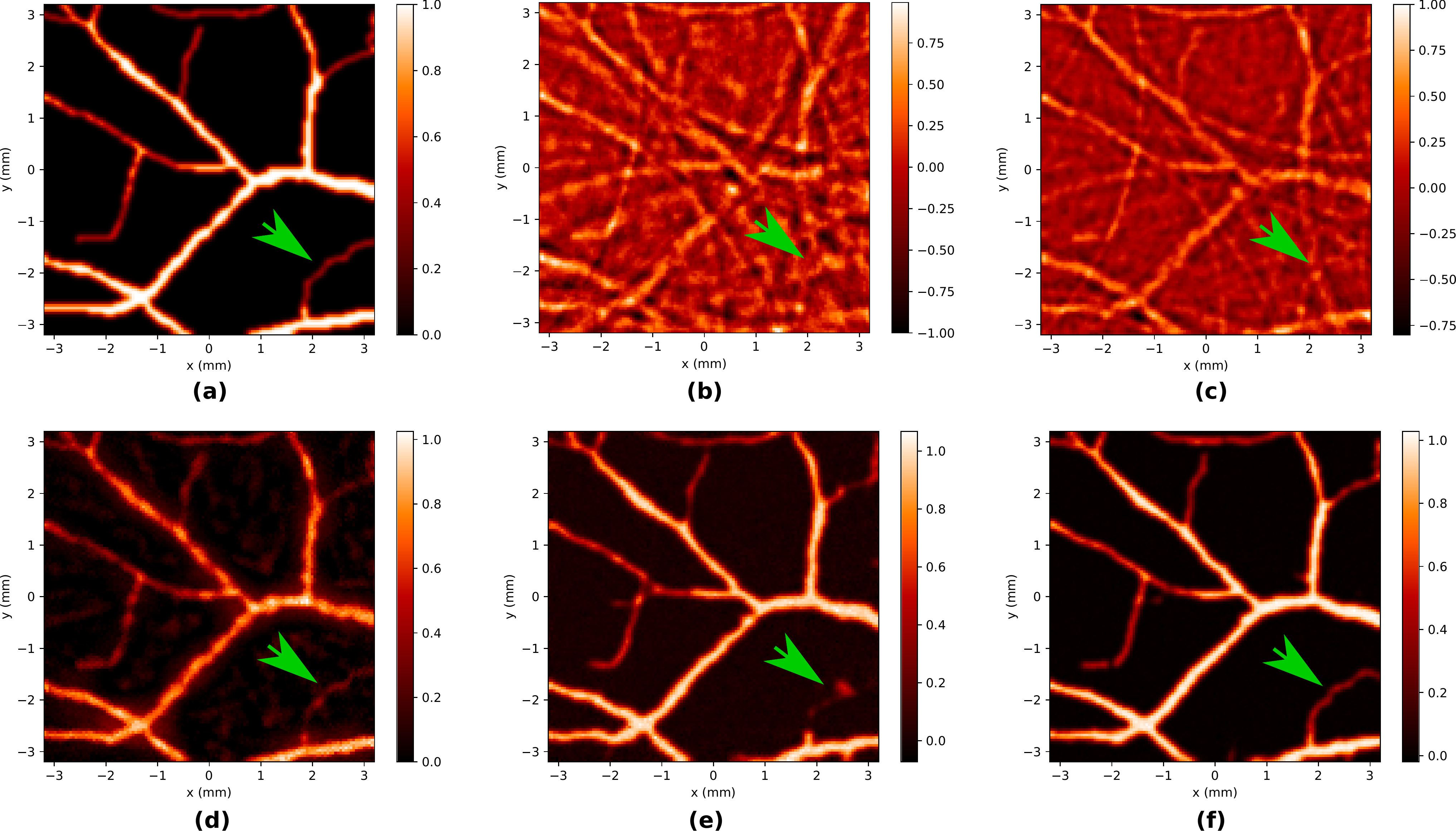}
    \caption{Reconstruction example for different methods: (a) true image; (b) DAS; (c) LBP; (d) fbMB; (e) FD-UNet; (f) ours (fb-FD-UNet). The green arrows indicate a small structure enhanced with the proposed method.}
\label{fig:1}
\end{figure}

\begin{figure}
  \centering
  \includegraphics[width=13.5cm]{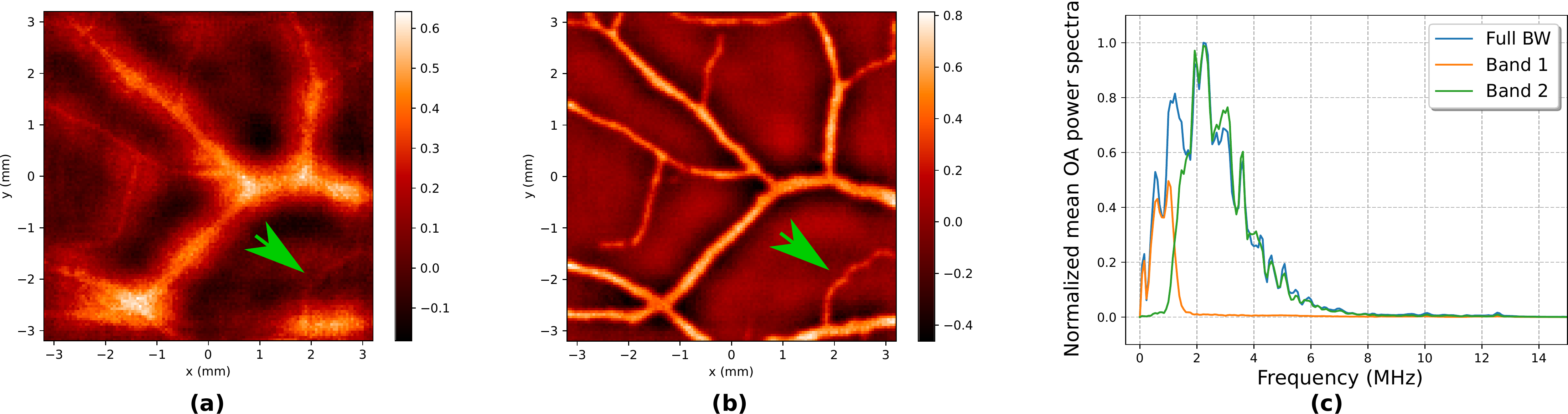}
    \caption{(a)-(b) Low and high-frequency image components for the fb-FD-UNet method. (c) The average power spectrum of the full sinogram and the components associated with $\mathbf{x}_1$ and $\mathbf{x}_2$.}
\label{fig:2}
\end{figure}

In Fig. \ref{fig:1} we show the qualitative performance for an image in the testing set for all considered methods. The considered image is a retinal vasculature phantom (subfigure (a)) that presents a very thin vessel indicated by the green arrow. It is expected that the information of this thin structure should be in the high-frequency content of the sinogram. In (b) and (c), we see the reconstructed images by DAS and LBP. Not only the small vessel is not clearly visible, but the whole reconstructed image has a low visual quality. In (d) and (e) we see the reconstruction achieved by fbMB and FD-UNet, respectively. General visual quality is significantly better, but the small vessel reconstruction is not good enough. In (e) we see the result of our architecture, where not only the visual quality of the whole image is the best, but also the small vessel can be easily spotted. In Fig. \ref{fig:2}, the image components obtained by our method are presented, showing the enhancement of the thin vessel in the high-frequency image $\mathbf{x}_2$. Also, the average power spectrum of the sinogram and its components associated with images $\mathbf{x}_1$ and $\mathbf{x}_2$ is depicted, showing clearly the frequency separation achieved by the second term in the loss function (\ref{eq:loss_2}). The sharp separation between the components, that it is observed in Fig. \ref{fig:2}(c), is a consequence of the zero-phase filtering considered in the construction of the band-reject matrices $\mathbf{F}_k$, $k=1,2$. 

One important issue we want to emphasize is the following. Equation \ref{eq:loss_1}, impose a non-negativity constraint in each of reconstructions $\mathbf{x}_k^*$ with $k=1,\dots,n$. This has the result of avoiding negative valued pixels at those images as it is usual in OA image processing. Visually, for the naked human eye, this has the effect of a more clear frequency separation in the multilevel images (but obviously not for the final full image). For the loss function \ref{eq:loss_2} corresponding to our method fb-FD-UNet, we have not imposed such a constraint. The first motivation to do this is the fact that, in principle, we are not necessarily interested in each of the images $\mathbf{x}_k^*$ with $k=1,\dots,n$ but in the quality of the final image $\hat{\mathbf{p}}_0=\sum_{k=1}^n \mathbf{x}_k^{*}$. Secondly, the imposing of non-negativity constraint requires either a modification on the loss function to enforce such non-negativity constraint or a modification of the final layer in the neural network to impose such constraint. We think that not forcing non-negativity constraints on $\mathbf{x}_k^*$ with $k=1,\dots,n$, allows to finding better parameters for the network when the final goal is good reconstruction of the whole image. For example, in Fig. \ref{fig:1}(f) we see the good quality reconstruction of our proposal where all the pixels are positive as the color reference at the right of image shows. On the other hand, in Figures \ref{fig:2}(a) and \ref{fig:2}(b), the multilevel images of our proposal are presented. We clearly see that the pixels in those images can be negative. In the low and high-pass images, this manifests in some blurring (more noticeably in the low-pass image) that disappears after linear aggregation (Fig. \ref{fig:1}(f)). This scheme provides the best quality reconstruction of all methods, even the one of \cite{Longo_Justel_Ntziachristos_2022} which, as explained above, enforces the non-negativity of the low and high-pass images. In order to check the performance of our method when a non-negativity constraint is imposed in the components $\mathbf{x}_k^*$ with $k=1,\dots,n$, we have slightly modified the architecture, including a ReLU layer (Rectified Linear Unit) at each of the output channels of the network. This inclusion has the net effect of guaranteeing the non-negativity of all component images $\mathbf{x}_k^{*}$. The architecture was then optimized using the training sequence in the same manner as done above selecting the best hyperparameters. Table \ref{table:1_aux} presents the results over the testing set for the four metrics. It can be appreciated that there is a degradation in the average behaviour in almost all metrics. Although in this case the multilevel images presents non-negative pixels (as in \ref{eq:loss_1}), the final performance is not better than the case in which there is no such constraint. In summary, in our proposed method, although each multilevel image $\mathbf{x}^{*}_k$ can have negative pixels, the network parameters are such that the final aggregation produces a reconstructed image $\hat{\mathbf{p}}_0$ which has non-negative ones.

\begin{table}
    \centering \caption{Performance over the testing set of our proposal with a non-negativity enforcement in component images}
    \begin{tabular}{cccc}
        \hline
        SSIM & PC & RMSE & PSNR \\
        \hline
     0.756 $\pm$ 0.125 & 0.964 $\pm$ 0.039 & 0.054 $\pm$ 0.021 & 25.966 $\pm$ 3.054 \\
        \hline
   \end{tabular}
\label{table:1_aux}
\end{table}

\begin{figure}[b]
  \centering
  \includegraphics[width=12.0cm]{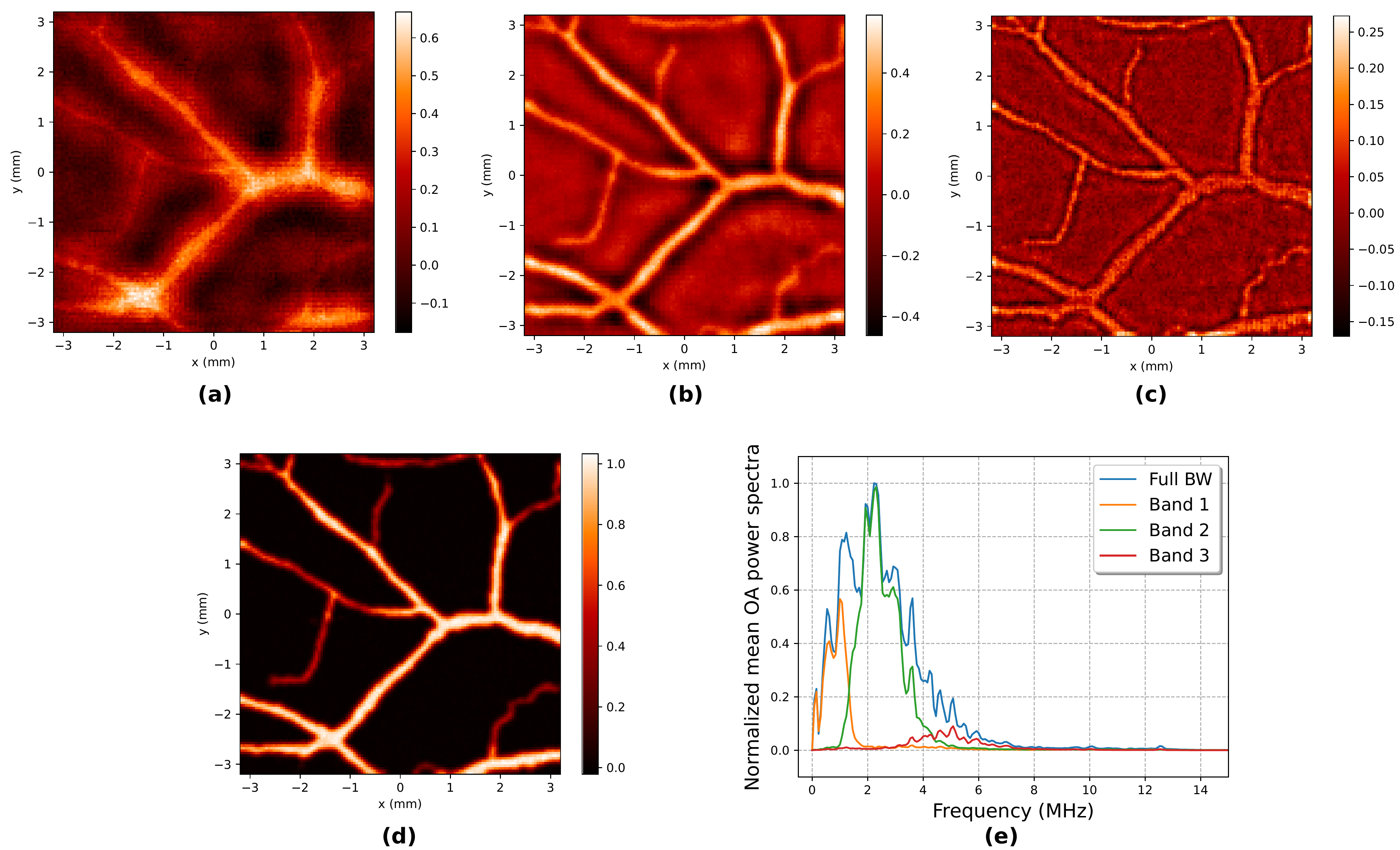}
    \caption{Reconstruction example of the proposed method using $n=3$. (a)-(c) Low, middle and high-frequency image components. (d) Predicted image. (e) The average power spectrum of the full sinogram and the components associated with $\mathbf{x}_1$, $\mathbf{x}_2$ and $\mathbf{x}_3$.}
\label{fig:3}
\end{figure}

We also considered the implementation of our method for $n=3$ to study if any significant gain is obtained from further frequency separation. In this case the frequency separation was also performed with Butterworth filters a common $BW_{\%}=1$ and with frequency bands of $[0.44 \mbox{MHz}, 1.33 \mbox{MHz}]$ (resolution up to 1.1 mm), $[1.33 \mbox{MHz}, 4 \mbox{MHz}]$ (resolution up to 0.37 mm) and  $[4 \mbox{MHz}, 12 \mbox{MHz}]$ (resolution up to 0.12 mm). In Table \ref{table:2} the performance over the testing set is presented. We see that the performance is similar to the $n=2$ case. Similarly, in Fig. \ref{fig:3} the same reconstruction example used for the $n=2$ case is shown. We observe again a similar performance. Moreover, we see that the high-frequency image corresponds to signal with small contribution in terms of mean power spectra (Fig. \ref{fig:3}(e)), but that at the same time presents a good visualization of the small vessel in the bottom right corner (Fig. \ref{fig:3}(c)). This allow us to conclude that, for the images considered in this work, there is not a significant improvement with respect to the $n=2$ case. However, for other images with finer and smaller details, some improvement can be expected if the number of bands at their cut-off frequencies are chosen carefully.

An important consideration, besides the performance on the image reconstruction, is the computational effort needed by each reconstruction method. Clearly, the computational cost of training the FD-UNet and our proposal is significant. However, once the network parameters are tuned, the computational complexity for image reconstruction is low. In order to check this for the experiment with $n=2$, we considered the use of only CPU instructions for evaluating each of the methods studied in this work. We registered the average times (over the testing set) for the four reconstruction algorithms. DAS and LBP are the most efficient, requiring only 0.03 and 0.07 sec. to process the sinogram and obtain the image. On the other end, fbMB required 255 sec. to deliver the reconstructed image. Our proposal, on the other hand, required only 0.18 sec. to process the sinogram, providing an excellent trade-off between reconstruction quality and computational effort.

\begin{table}[h]
    \centering \caption{Performance over the testing set of our proposal with $n=3$}
    \begin{tabular}{cccc}
        \hline
        SSIM & PC & RMSE & PSNR \\
        \hline
     0.880 $\pm$ 0.104 & 0.965 $\pm$ 0.043 & 0.047 $\pm$ 0.022 & 27.529 $\pm$ 4.263 \\
        \hline
   \end{tabular}
\label{table:2}
\end{table}

\begin{figure}[b]
  \centering
  \includegraphics[width=11.5cm]{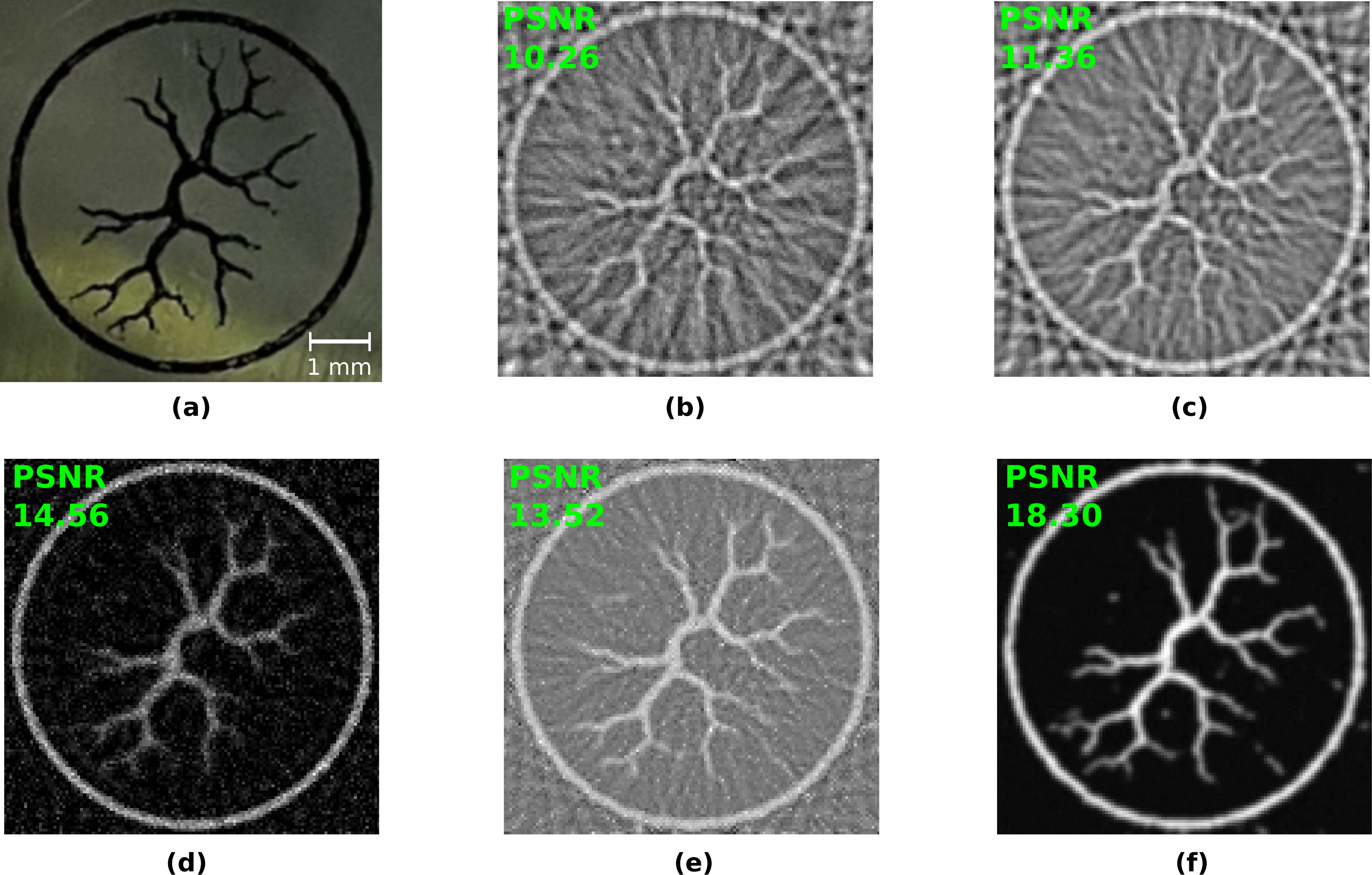}
    \caption{Reconstruction example using experimental measurements for different methods and their respective PSNR: (a) true image; (b) DAS; (c) LBP; (d) fbMB; (e) FD-UNet; (f) ours (fb-FD-UNet). 
    }
\label{fig:expresults}
\end{figure}

Finally, in order to test the performance of our method under experimental conditions, we used the two-dimensional OAT system described in \cite{Hirsch_Gonzalez_ReyVega_2021}. The sample consists of an ink pattern laser (artificial vein image) printed on a transparent film embedded in agarose gel.  A picture of the sample is shown in Fig. \ref{fig:expresults} (a). The OA signals were acquired over $N_d = 32$ locations placed equidistantly around a circumference, recorded ($N_t = 1024$, $\Delta t = 12$ ns) with a SNR of 26 dB.  The distance between the sensor and the center of the rotating sample was $8.5$ mm with an estimated position uncertainty of about $0.1\%$. The speed of sound of the medium between the sample and the detector was  $v_s=1485$ m/s. The reconstructed images are shown in Fig. \ref{fig:expresults}.

It can be seen that our proposal has the best qualitative result. It is also observed that fbMB also obtains an acceptable reconstruction quality. Moreover, the performance in terms of PSNR is shown for each method. It can be appreciated that our proposal and fbMB methods give the best results. However, our approach provides a gain of almost 4 dB with respect to fbMB. In this example, the simple FD-UNet architecture performs worse than fbMB, although only a difference of 1 dB is observed which is a value within the standard deviation reported in Table \ref{table:1}.

\section{Conclusions}
\label{sec:conclu}

We considered an image reconstruction algorithm for the application of OAT. Our proposal make use of the frequency disentanglement idea considered in \cite{Longo_Justel_Ntziachristos_2022}, an appropriate neural network architecture (FD-UNet) and a careful chosen loss function that simultaneously exploit the presence of ground-truth images in the training dataset, the data consistency between the sinogram and the reconstructed image and separation in frequency-bands of the sinogram. Numerical experiments, using public databases and real measurements, show that the proposal method is competitive with respect to several well-known quality metrics for the OAT image reconstruction problem. It also presents (after the more demanding training phase) computational advantages that can be relevant for real-time implementations in standard hardware. With respect to the real measurements results, although the ground truth image used in Fig. \ref{fig:expresults} is not as complex as the ones used in \cite{Longo_Justel_Ntziachristos_2022}, the value of these experimental results is to show that our reconstruction method, not only works comparatively better that other approaches for a database as \cite{drive2020} but also performs better in an experimental setting that not necessarily have the same exact characteristics that the one used to train our model. 

Although we have explored the possibility of using more than $n=2$ frequency bands and conclude that for specific images considered in this paper there is not a significant gain for $n=3$, further studies should be carried on.

Finally, it is important to highlight that, in this work, we have focused in the reconstruction quality of the final image which is the simple sum aggregation of the each the multilevel images $\mathbf{x}_k^*$ with $k=1,\dots,n$ and not on the individual quality of each of the multilevel images. It is clear that those images could have some value for identification and diagnosis and further work in this respect could be done in the future. 

\section* {Acknowledgments}
This work was supported by the University of Buenos Aires (grant UBACYT 20020190100032BA), CONICET (grant PIP 11220200101826CO) and the ANPCyT (grants PICT 2018-04589, PICT 2020-01336).

\section* {Author's contributions}
All authors contributed equally to this work.

\section* {Data Availability Statement}
The data that support the findings of this study are available from the corresponding author upon reasonable request.

\section* {Declaration of competing interest}
The authors declare that they have no known competing financial interests or personal relationships that could have appeared to influence the work reported in this paper.

\bibliographystyle{apalike}
\bibliography{references}  

\end{document}